\documentstyle[12pt]{article}

\def\eop{\vspace*{\fill}\pagebreak}

\begin{document}

\begin{titlepage}
{\bf June  1995}\hfill       {\bf PUPT-1546}\\
\begin{center}

{\bf TURBULENCE WITHOUT PRESSURE}

\vspace{1.5cm}

{\bf  A.M.~Polyakov}

\vspace{1.0cm}

{\it  Physics Department, Princeton University,\\
Jadwin Hall, Princeton, NJ 08544-1000.\\
E-mail: polyakov@puhep1.princeton.edu}

\vspace{1.9cm}
\end{center}

\abstract{
We develop exact field theoretic methods to treat turbulence  when the effect
of pressure is negligible. We find explicit forms of certain probability
distributions, demonstrate that the breakdown of  Galilean invariance is
responsible for intermittency and establish the operator product expansion. We
also indicate how the effects of pressure can be turned on perturbatively.
}
\vfill
\end{titlepage}

Turbulence is an old and tantalizing subject. Enormous amounts of data  and
ideas have been accumulated during  this  century and still the problem is not
solved. In our opinion, the reason lies in the fact that  the necessary
field-theoretic tools have appeared only recently. \par Two years ago an
attempt was  made to apply the methods of conformal field theory to the case of
two-dimensional turbulence [1]. The main concepts of this work were the
following.
First, one looks at the steady state condition, which relates the equal time,
N- point and N+1 - point functions. Then one argues that in the inertial range
these relations can be solved exactly by  field theories satisfying fusion
rules or operator product expansions (OPE). There appeared to be infinitely
many solutions. An additional constraint on these solutions follows from the
constant flux conditions.\par  It has been noticed (although not really
exploited) in [1] that there exists a striking analogy between the constant
flux states in turbulence and axial (and other) anomalies in quantum field
theory. The latter are violations of the naive conservation laws caused by the
ultra-violet regularization. In the case of turbulence the  ultraviolet
regularization arises from viscosity and results in an energy flux through the
inertial range.
When  the steady state condition with these two tools was analysed it appeared
that the third ingredient was needed. Namely the physical correlation functions
contained  so-called condensate terms, which were    $ \delta$ - functions in
the momentum space and represented the large scale motions of the fluid.  Their
role was to cancel infrared divergencies yhat arose from the field-theoretic
fluctuations.
Precise form of these terms depends on the large scale region were the energy
is pumped into the system. The task of joining the  inertial range with this
region remained unsolved in [1]. \par
It is highly desirable to have an exactly soluble model in which the above
ideas can be tested at work. In this article we will discuss such a model,
which also is of independent physical interest.
The model in question is simply the Navier-Stokes equation with white noise
random force and with the  pressure set equal to zero. In one dimension this is
known as the  Burgers equations.
\par Such equations have been exploited in the past in many different physical
situations (like galaxy formations [ 2 ], crystal growth [ 3 ] etc). Recently
they were the subject of deep mathematical investigations [ 4 ] \par In a
remarkable paper  [ 5] some striking numerical data concerning Burgers
turbulence were obtained and an appealing qualitative  picture of the
phenomenon has been proposed. This work to a  large extent inspired my interest
in turbulence without pressure. Another important work in this area is the
recent paper [6] on which I shall comment later.
\par In  the present  paper we shall formulate a general new method for
analyzing the inertial range correlation functions, based on the ingredients
mentioned above (OPE and anomalies). The method with minor modifications is
also applicable to the problem of advection of passive scalars and other cases.
It is obvious that the ideas we develop below will become a part of the general
theory of turbulence. They may also have a considerable "back reaction" on the
field theory.
\par Let us start with the  one dimensional case.  Burgers equation has the
form:
\begin{eqnarray}
u_{t}+uu_{x}=\nu u_{xx} +f(xt) \\
< f(x ,t) f(\acute{x} ,\acute{t}) =\kappa (x-\acute{x}) \delta (t- \acute{t})
\nonumber
\end{eqnarray}
Here the function $\kappa$ defines the spatial correlation of  the random
forces.
 Consider the following generating functional:
\begin{equation}
Z( \lambda_{1} x_{1} | \ldots | \lambda_{N} x_{N} ) = \left\langle
\exp{\sum{\lambda_{j}u(x_{j} t)}} \right\rangle
\end{equation}

{}From (1) we derive :

\begin{equation}
\dot{Z}+ \sum{\lambda_j{\partial \over \partial \lambda_j}({1 \over
\lambda_j}{\partial Z\over \partial x_j})} = \sum\lambda_j\left\langle
(f(x_j t)  +\nu u'' )\exp{\sum{\lambda_{k}u(x_{k}t)}} \right\rangle
\nonumber\end{equation}
The first term in the right hand side is easy to treat since the force $ f(x t)
$ is Gaussian and we can use the standard trick [ 7] of the theory of Langevin
equations:
\begin{equation}
\left\langle f(xt)\exp{\sum{\lambda_{j}u(x_{j}t)}} \right\rangle =\sum{\kappa
(x-x_{j})\lambda_{j}} Z
\end{equation}
Our equation takes the form:
\begin{eqnarray}
\dot{Z} + \sum\lambda_{j}{\partial \over \partial \lambda_{j}}({1 \over
\lambda_{j}}{\partial {Z}\over \partial x_{j}}) = \sum{\kappa (x_{i} - x_{j})
\lambda_{i} \lambda_{j}} Z + D
\end{eqnarray}
By D we denote the dissipation term:

\begin{eqnarray}
D = \nu \sum\lambda_{j} \left\langle u'' (x_{j} t) exp{\sum{\lambda_{k}
u(x_{k}t)}} \right\rangle
\end{eqnarray}
   If the viscosity $ \nu $  were zero our task would be completed since we
have a  closed differential equation for Z. To  reach the inertial range we
must, however, keep $\nu $ infinitesimal but non-zero. The anomaly mechanism
mentioned above implies that infinitesimal viscosity produces a   finite
effect, whose computation is one of our main objectives. First, however, let us
transform and interpret the  inviscid equations (5)  ( dropping the D-term).
\par Let us introduce the function $ F$  given by :

\begin{eqnarray}
Z = \lambda_{1} \ldots \lambda_{N} F(\lambda_{1}  x_{1}\ldots \lambda_{N}
x_{N})
\end{eqnarray}

We have:
\begin{eqnarray}
\dot{F} +\sum{\partial^{2} \over \partial  x_{k}\partial \lambda_{k}} F -
\sum{\kappa (x_{i} - x_{j})\lambda_{i} \lambda_{j}}F = \tilde{D}
\end{eqnarray}
Here:
$$\tilde{D}={D \over \prod{\lambda_{j}}}$$
We can now introduce the  Fourier- transform  ${\cal F} ={\cal F} (u_{1}x_{1}
\ldots u_{N} x_{N})$

which satisfies:
\begin{eqnarray}
\dot{\cal F} +\sum u_{k} {\partial \over \partial x_{k}}{\cal F}- \sum\kappa
(x_{i}-x_{j}){\partial^{2} \over \partial u_{i} \partial u_{j}}{\cal F} =
\tilde{\cal{D}}
\end{eqnarray}

obtained from(8) by the  substitution
  $ \lambda \Rightarrow  {\partial \over \partial u}$

The function $ \cal F $  has a  simple meaning. It can be interpreted as:
\begin{eqnarray}
{\cal F }=\left\langle \theta (u_{1}x_{1}) \ldots \theta (u_{N}x_{N})
\right\rangle
\end{eqnarray}

where $ \theta (u_{k} x_{k})= \theta (u_{k} - u(x_{k}t)) $

and  the last $ \theta$  being a step function. In order  to get the N-point
probability distribution $Z $ one has to differentiate $ \cal F$ according to
(7):

\begin{eqnarray}
Z_{N}={\partial^{N} \over \partial u_{1}\ldots \partial u_{N}} {\cal F}_{N}
\end{eqnarray}
We supplied here our correlation functions with the subscript  $N$ to indicate
the number of points on which these functions depend.
\par Of course, equation (9) could have been obtained directly by computing
time derivative of the $ \theta$ - field. It is also easy to write to express
the $D_{N}$ - term through $ {\cal F}_{N+1}$
By using (6) we obtain:

$$D_{N}=\sum{{\partial \Gamma^{(N)}_{j} \over \partial u_{j}}} $$
\begin{eqnarray}\Gamma^{(N)}_{j} = \nu {\partial^{2} \over \partial
y^{2}}{\int{u_{0}du_{0} {\partial \over \partial u_{0}}{{\cal F}_{N+1}(u_{0},
x_{j}+y ; u_{1},x_{1} \ldots u_{N},x_{N})}}} \mid _{y\rightarrow 0}
\end{eqnarray}

Equations (9) and (12) give a chain of relations remarkably similar to the
BBGKY equations of  statistical mechanics [7]. One can hardly hope to solve
these equations exactly. But we are interested in the inertial range, which
means that we have to take the limit $ \nu\rightarrow 0$. We will show now that
in this case the system of equations closes and gives us an equation for
turbulent kinetics, much in a same way as the Boltzman equation becomes exact
in the limit of small densities.
\par The main ideas of the  derivation are the following. For the large
Reynolds numbers, corresponding to small $ \nu  $ there are two relevant
scales. The first, $L$ , is  defined by the size of the system  and provides an
infrared cut-off. The second ,$ a\ll L$,  is the scale at which dissipation
becomes relevant. The ratio  ${L \over a} $ goes to infinity together with the
Reynolds number. By the existence of the inertial range we mean the conjecture
that the correlation functions $ \cal F$   have a  finite limit at zero
viscosity, provided that we keep $ x_{i}-x_{j} $ fixed.
They can have singularities at coinciding points, which must be understood as
being   smeared by the  viscosity at the scale $ a$ . In fact, this scale is
determined by the condition that as we let  $ \nu $ go to zero, dissipation
remains finite.
 This means, that we have to find the leading singularity in (12)  as $
y\rightarrow 0$, and compensate for it by an  appropriate scaling of $\nu (y) $
. All subleading terms will give vanishing contribution in the inertial range
(in the limit of the infinite Reynolds numbers).
The task of finding the leading singularities is precisely what  the OPE was
developed for.
\par However, we should warn the reader that what follows is essentially a self
- consistent conjecture.  In the case of statistical mechanics, when deriving
hydrodynamics from the BBGKY equations it is necessary to assume the decrease
of the correlations [7 ], a self-consistent assumption that is  difficult to
prove from  first principles. In our case this property is replaced by the OPE.
\par To understand  how they work, let us reexamine the derivation of the
previous equations. They were based on the fact  that modulo the stirring force
and the  viscosity we have a sequence of conservation laws:

\begin{eqnarray}
{\partial \over \partial t}(u^{n}) +{n \over n+1}{\partial \over \partial
x}(u^{n+1})\approx 0
\end{eqnarray}
(The sign $\approx$ here means that we don't write terms coming from the
viscosity and the stirring force)
\par Equations (5) and (9) can be interpreted as relations for the generating
functionals $ \left\langle u^{n_{1}}(x_{1})\ldots u^{n_{k}}(x_{k})
\right\rangle $. They involve both the stirring force and the viscosity. The
former was already accounted for, while the latter presents a problem. The main
rule of the game is that in any equation ,involving  space points separated by
the distance larger than $a$ viscosity can be set to zero. Thus it is perfectly
legitimate to use the inviscid limit for n=1, because in this case we exploit
the  steady state condition:
\begin{eqnarray}
{d \over dt}{\left\langle u(x_{1}) \ldots u(x_{N}) \right\rangle}=0 \nonumber
\\ \mid x_{i}-x_{j}\mid \gg a
\end{eqnarray}
However, starting from n=2 we have a problem, since in this case we have to
take time derivatives of the product of $ u$  - s at the same point. To
circumvent this problem in the case of  n=2  let us  replace:
\begin{eqnarray}
u^{2}(x)\Longrightarrow u(x+{y \over 2})u(x-{y \over 2})\nonumber \\ \mid
x_{i}-x_{j}\mid \gg y\gg a
\end{eqnarray}
 and let $ y\rightarrow 0$ after the viscosity is taken to zero. In this case
the use of the inviscid equations is justified, but we will get an anomaly in
the conservation law, due to the point splitting. We have  after simple
algebra:
\begin{eqnarray}
-{d \over dt}{(u(x+{y \over 2})u(x-{y \over 2}))}\approx {1 \over 2}{\partial
\over \partial x_{1}}{(u^{2}(x_{1})u(x_{2}}))+(1\leftrightarrow 2) \approx{2
\over3}{\partial \over \partial x}{u^{3}(x)}+ a_{0}(x),
\end{eqnarray}
$$ x_{1,2}=x\pm {y \over 2}$$
Here we have  introduced the first dissipative anomaly operator:
\begin{eqnarray}
a_{0}(x)=\lim_{y\rightarrow 0}{{1 \over 6}{\partial \over \partial y}{(u(x+{1
\over 2}y)-u(x-{1 \over 2}y))^{3}}}
\end{eqnarray}
In deriving this formula we set $y$ to zero  inside all terms containing $x$-
derivative. This is possible because all the  correlation functions have finite
 $x$ -dependent limit  at zero $y$. We also used  the identity:
$$ {\partial \over \partial y}{(u^{3}(x+{y \over 2})-u^{3}(x-{y \over 2}))}={1
\over 2}{\partial \over \partial x}{(u^{3}(x+{y \over 2})+u^{3}(x-{y \over
2}))} \rightarrow {\partial \over \partial x}{(u^{3}(x))}$$
The anomaly  would be zero if $ u(x)$ were differentiable, since then the RHS
of (17) is $\sim y^{2}$. However, the steady state condition dictates the
opposite. Indeed, one of the consequences of  eq. (5) is that in the steady
state:
\begin{eqnarray}
{d \over dt}{\left\langle u^{2} \right\rangle}=\kappa (0) -\left\langle a_{0}
\right\rangle =0
\end{eqnarray}
and hence we have the famous Kolmogorov relation:
\begin{eqnarray}
\left\langle (u(x_{1})- u(x_{2}))^{3} \right\rangle \propto \kappa
(0)(x_{1}-x_{2})
\end{eqnarray}
The value of the anomaly defines the limiting contribution of the viscous term
in the steady state:
$$ \lim_{\nu\rightarrow 0}{\nu u(x)u''(x)} = -a_{0}(x)$$
An interesting feature of this relation is that it defines the expectation
value of the $a_{0} $ - anomaly self-consistently from the steady state
equation. This feature is preserved for the  higher anomalies of  the $ u^{n}$
- densities. They are necessarily non-zero, because after point splitting
procedure we get terms$\sim {\partial \over \partial y}{(u(x+{y \over
2})-u(x-{y \over 2}))^{k}} $
 and the steady state equation will determine their value.\par
With a certain amount of vulgarization one can say that the reason for the $
u^{n}$ anomalies is that  shock waves absorb not only energy, but these higher
densities as well. \par Before computing the general anomaly let us discuss
carefully all the  limiting procedures involved. As we see correlation
functions depend on the parameters $\kappa (0)=\varepsilon $, $L$, (which
defines the correlation length of the forces)and the viscosity $\nu$. We made
an assumption  that  as we let the viscosity  go to  zero correlation functions
have  a finite limit and hence depend only  on $\varepsilon $ and $ L $ . This
limit is what is meant by the inertial range. The standard Kolmogorov
assumption
(which we don't make) is that the Galilean - invariant correlation functions,
such as  $\left\langle (u(x_{1})-u(x_{2}))^{n}\right\rangle $ have a finite
limit as $ L\rightarrow \infty $. As we will see in our case this statement
doesn't hold. Instead we have to make a different assumption, consistent with
our equations. We will call it the  G- (Galilean) assumption. \par To formulate
it let us notice that  Galilean invariance in our system is spontaneously
broken. This is evident from the fact that pumping forces create a certain
average  velocity $ v_{rms} = \surd \left\langle u^{2} \right\rangle $. At the
same time  unbroken G-symmetry would require that  the probability distribution
  be invariant under  $u\Rightarrow u+const $. \par It is easy to estimate the
value of $v_{rms}$. In the limit of zero viscosity the only possible formula is
:
\begin {eqnarray}
v_{rms}\sim \kappa^{{1 \over 3}}(0)L^{{1 \over 3}}\sim \varepsilon^{1 \over
3}L^{{1\over 3}}
\end{eqnarray}
In the $\lambda $ representation this breakdown means that:
\begin{eqnarray}
\sum{\lambda_{j}}\sim {1 \over v_{rms}}
\end{eqnarray}
This absence of G-symmetry makes the  anomaly computation difficult.
Fortunately it is consistent with eq.(3) to assume that if we formally tend $
L\rightarrow \infty $ and keep $ \lambda_{j} $ finite, then G-symmetry is
restored. We  conjecture that in this limit:
\begin{eqnarray}
Z(\lambda_{1}x_{1};\ldots \lambda_{N}x_{N}) \propto \delta(\sum{\lambda_{k}})
\end{eqnarray}
This is the  G-assumption and a short check shows that it is  consistent with
eq.(3). To state this assumption in a slightly more physical way, we can say
that the  probability  distributions of velocities, $ W(u_{1}x_{1} \ldots
u_{N}x_{N}) $, which are Fourier transforms of the $Z$ functions, have the
following structure:
\begin{eqnarray}
W(u_{1}x_{1} \ldots u_{N}x_{N}) = w(u_{i}-u_{j} ; x_{k}) W_{1}\left
({\sum u_{k}} \over {v_{rms}}\right )
\end{eqnarray}
provided that :
\begin{eqnarray}
\mid u_{i}-u_{j}\mid \ll \sum{u_{k}}
\end{eqnarray}
This last condition is very important. It is easy to see  that without it  the
separation of the center of mass velocity which occured in (23) would
contradict eq.(5). \par The G-symmetry greatly simplifies computations of the
anomaly. However one more self-consistent assumption is needed . This is an
assumption of the existenceof an  operator product expansion or the fusion
rules. To formulate it we introduce  the following notations:
\begin{eqnarray}
Z(\ldots)=\left\langle e_{\lambda _{1}} (x_{1}) \ldots e_{\lambda_{N}} (x_{N})
\right\rangle \nonumber \\ e_{\lambda} (x)=\exp{\lambda u(x)}
\end{eqnarray}
The fusion rules is the statement concerning the behavior of  correlation
functions when  some subset of points  are  put close together. We conjecture
that in our case  the rules  have the form:
\begin{eqnarray}
e_{\lambda_{1}} (x+{y \over 2}) e_{\lambda_{2}} (x-{y \over 2}) =
A(\lambda_{1}, \lambda_{2}, y) e_{\lambda_{1}+\lambda_{2}}(x) + B(\lambda_{1},
\lambda_{2}, y) {\partial \over \partial x}{e_{\lambda_{1}+\lambda_{2}}} +
O(y^{2})
\end{eqnarray}
We will call this statement the F-conjecture. Here  $ A $ and $ B $ are some
functions to be determined  and the meaning of  (26) is that they control the
fusion of the functions $ Z_{N}$ into functions $ Z_{N-1} $ as we fuse a couple
of points together. To find the result one has to substitute (26) into (25). Of
course one must check that this conjecture is consistent with the eq. (5),
which is also is supposed to determine functions $A$ and $B$. To make this
equation effective we have to evaluate the following anomaly operator:
\begin{eqnarray}
a_{\lambda }(x) = \lim_{\nu \rightarrow 0}{\nu (\lambda u''(x) \exp{\lambda
u(x)})}
\end{eqnarray}
which appears on the right hand side. As we explained above, $ a_{\lambda}(x) $
is generally non-zero, because smallness of $ \nu $ is compensated by the
blow-up of  $ u''(x+y)\exp{\lambda u(x)}$ as $ y\rightarrow 0 $. In fact we can
write:

\begin{eqnarray}
a_{\lambda }(x) =\lambda \lim_{\nu , y, \zeta\rightarrow 0}{\partial^{3} \over
\partial \zeta \partial y^{2}}{e_{\zeta }(x+y) e_{\lambda }(x)}
\end{eqnarray}
 and exploit the F-conjecture to evaluate the  RHS of  (28).
Thus,  $a_{\lambda }(x) $ should be expressed in terms of derivatives of the
functions $ A $ and $ B $. If the result is finite it must have the form:

\begin{eqnarray}
a_{\lambda }(x) = \alpha (\lambda )e_{\lambda }(x) +\tilde{\beta}(\lambda
){\partial \over \partial x}{e_{\lambda }(x)}
\end{eqnarray}
which is the only possible G-invariant expression,involving  the ultraviolet
finite operators $e(x) $ and $ u'(x) $. In order to have a  finite limit in
(28) one has to set the cut- off values of $ y $ and $ \zeta $ to be  dependent
on $\nu$ and $\lambda $. This is not surprising, since ${1 \over \lambda }$
defines the local Reynolds number and the ultraviolet cut-off must depend on
it. It is also worth stressing again that this form of the anomaly is correct
only in the Galilean- invariant limit.  For  generic $ \lambda_{j} $ we would
obtain a superposition of exponents with different  $ \lambda$ , a rather
difficult  situation to treat. \par The master equation (5) now takes the form
(for the steady state):
\begin{eqnarray}
 HZ\equiv\sum ({\partial \over \partial \lambda_{j}}- \beta
(\lambda_{j})){{\partial \over \partial x_{j}}{Z}} -\sum{\kappa
(x_{i}-x_{j})\lambda_{i} \lambda_{j}}Z=\sum{\alpha (\lambda_{j})}Z
\end{eqnarray}
 $$\beta (\lambda )=\tilde{\beta}(\lambda) + {1 \over \lambda}$$

\par It is now a simple matter to check our F-conjecture. If we introduce the
variables:
\begin{eqnarray}
x_{1,2} =x \pm {y \over 2} \nonumber \\ \lambda =\lambda_{1} +\lambda_{2}
\nonumber \\ \mu =\lambda_{1} -\lambda_{2} \nonumber
\end{eqnarray}
and keep $ y $ much smaller than all other distances, we find the following
structure of the operator  $ H $ :
$$H_{N}=H_{N-1}+F$$
\begin{eqnarray}
F= 2{\partial^{2} \over \partial y \partial \mu} -(\beta (\lambda_{1}) - \beta
(\lambda_{2})){\partial \over \partial y}+(\beta (\lambda)-{1 \over 2}(\beta
(\lambda_{1})+\beta (\lambda_{2})){\partial \over \partial x}
\end{eqnarray}
Here  the operator $ H_{N-1} $  is obtained from $H_{N}$  by replacing  the
points $ x_{1,2} $  with the point $x$ and $ \lambda_{1,2} $ with $\lambda =
\lambda_{1} +\lambda_{2} $. From this we derive equations for the functions $A$
and $B$  that appear in (26):

\begin{eqnarray}
(\nabla_{1}-\nabla_{2}){\partial {B}\over \partial y}={1 \over
2}(\beta(\lambda_{1})+\beta (\lambda_{2})) -\beta(\lambda_{1} +\lambda_{2})
\nonumber \\ (\nabla_{1}-\nabla_{2}){\partial {A}\over \partial y} =\alpha
(\lambda_{1})+\alpha(\lambda_{2})-\alpha (\lambda_{1}+\lambda_{2}),
\end{eqnarray}
Here:
$$\nabla ={\partial \over \partial \lambda} -\beta (\lambda)$$

These equations have solutions for any functions $\alpha$ and $\beta$. The next
step is to substitute these solutions back in eqs. (28) and (29) and try to
find constraints on $\alpha$ and $\beta$. Surprisingly the arising constraints
are very weak due to the possibility to adjust the cut-off's, and thus the
functions  $\alpha$ and $ \beta$ in (29)  remain almost  arbitrary.
In principle they
must be determined from the conditions that all probability distributions have
admissible behaviour at  $\infty $, much in the same way in which eigenvalues
are usually determined. Since we don't have any general methods for treating
this problem we will simplify the matter even more  by introducing a  scaling
conjecture ( S-conjecture) which again turns out to be self- consistent. To
formulate it  let us notice that if $ x\ll L $ we can expand $\kappa
(x)=\kappa(0)(1-{x^{2} \over L^{2}}) $ and in (30) the constant part ,
$\kappa(0)$ drops out due to the G-invariance:
\begin{eqnarray}
\sum{\kappa(x_{i}-x_{j})\lambda_{i} \lambda_{j}}=\kappa
(0)(\sum{\lambda_{i}})^{2} + O(x^{2}) \nonumber \\ \sum{\lambda_{i}}=0
\end{eqnarray}
Therefore it is natural to look for a scaling solution with $ \lambda \sim {1
\over x}$. The scaling condition determines the possible form of the functions
$\alpha$ and $\beta$. In order to conform to scaling , they must be as
following:

$$\beta (\lambda) = {b \over \lambda}$$
$$\alpha (\lambda )= a$$

We will see now that scaling is self-consistent, although one can also try more
general solutions, say with logarithmic terms. \par Let us see  how the
unknown numbers $a$ and $b$ are determined from the eigenvalue problem. For
example consider a 2-point function. The master equation in this case  takes
the form:
\begin{eqnarray}
({\partial \over \partial \mu}-{2b \over \mu}){\partial \over \partial
y}{Z}+\mu ^{2} y^{2} Z = aZ
\end{eqnarray}
For  reasons to be clarified later we are interested in the case $a=0$. Our
S-conjecture amounts in the anzats:
\begin{eqnarray}
Z(\mu , y)=\Phi (\mu y)
\end{eqnarray}
Here we  temporarily use the units in which $\kappa (0) =1$ and $L=1$
The function $\Phi (x) $ satisfies an ordinary differential equation:
\begin{eqnarray}
x\Phi ''(x) +(1-2b)\Phi '(x) +x^{2}\Phi (x) =0
\end{eqnarray}
The general solution of this equation has the form:
\begin{eqnarray}
\Phi (x)=x^{b}F_{{2b \over 3}}({2 \over 3}x^{{3 \over 2}})
\end{eqnarray}
where $F_{\lambda}$ is one of the Bessel functions. The right function and the
value of $b$ are determined from the condition that the probability
distribution:
\begin{eqnarray}
w(u , y)=\int_{c-i\infty}^{c+i\infty}{{d\mu \over 2\pi i}Z(\mu ,y)e^{-\mu u}}
\end{eqnarray}
must be positive and  vanish as $ u\rightarrow \pm \infty $.   From the
convergence of (38) it follows that we must chose
\begin{eqnarray}
\Phi (x) \propto x^{b}K_{{2b \over 3}}(-{2 \over 3}x^{{3 \over 2}})
\end{eqnarray}
 Positivity  of $Z$  for $x>0$ and its finitness at $x=0$ forces us to take
$b={3 \over 4}$ for which
case:
\begin{eqnarray}
\Phi (x)=\exp{{2 \over 3}x^{{3 \over 2}}}
\end{eqnarray}
As a result we obtain the following result for the probability to have  a
velocity difference $u$ at  the distance $y$ :
\begin{eqnarray}
w(u,y)=\int_{c-i\infty}^{c+i\infty}{{d\mu \over 2\pi i}\exp{[{2 \over 3}(\mu
y)^{{3 \over 2}}-\mu u}]}
\end{eqnarray}
The positivity of $w$ is guaranteed by the fact that $Z$ satisfies certain
convexity conditions. Indeed, from the relation:
\begin{eqnarray*}
Z(\mu)=\int_{-\infty}^{\infty}{w(u)e^{\mu u}}>0
\end{eqnarray*}
we derive , by using Cauchy inequality:
\begin{eqnarray}
Z(\mu_{1}) Z(\mu_{2})>Z^{2}({\mu_{1}+\mu_{2} \over 2})
\end{eqnarray}
which is  clearly satisfied. It is mathematically curious that the  $u$
representation  the eq.(34) is easily reduced to the Schroedinger equation with
the potential  $ V(u) ={u^{4} \over 4}-2bu$. This is achieved by the change:
$$ Z(\mu , y)=(\mu )^{2b}F(\mu, y) $$,
which removes the ${1 \over \mu}$ term in the eq. (34) and by the  Fourier
transform to the function  $\tilde{F}(u,y)=y^{2b-1}\phi ({u \over y})$. The
power of $y$ here is needed for consistency with (35).
  The value of $b$ corresponds to
the zero energy eigenvalue in this potential and $\phi$ is proportional to the
ground state wave function \footnote{A. Migdal informed me that he was able to
find this wave function directly from the  Schroedinger equation}. \par The
probability distribution defined by (41)has the following  asymptotic
behaviour:
\begin{equation}
 w(u,y) = \left \{ \begin{array} {ll} e^{-{1 \over 3}({u \over y})^{3}}&
\mbox{if ${u \over y}\rightarrow +\infty$ }\\ y^{{3 \over 2}}u^{-{5 \over 2}}
&\mbox{if ${u \over y}\rightarrow -\infty$}
\end{array} \right.
\end{equation}
This qualitatively fits the observations [5 ] . It must be stressed however
that some caution is needed when comparing  the G-invariant part of the
probability distributions with the experiment. As we  have already said, the
factorization (24) breaks down at large velocities. That means in particular
that  in general $w(u,y)$  has the following structure:
\begin{eqnarray*}
w(u,y)=\chi ({uL \over v_{rms}y}; {u \over v_{rms}}; {y \over L})
\end{eqnarray*}
The scaling limit, discussed above is reached only when two conditions are
satisfied:
\begin{eqnarray*}
y\ll L  ;  u\ll v_{rms}
\end{eqnarray*}
When  computing the moments of the probability distributions, which represent
correlation functions, it  may happen that even when the first condition is
enforced, the second one will be violated. In this case the result  is not
universal, since the behaviour of  probabilities at $u \sim v_{rms} $ depends
on the correlations of the stirring forces at $ x\sim L $.  \par We come to the
conclusion that the breakdown of Galilean invariance leads to a rather peculiar
structure of the correlation functions. They contain in general both universal
and non-universal parts. The former comes from the distribution (43) and its
generalization for an arbitrary number of points. The latter results from the
region of  the large velocities.  These  nonuniversal correlations are  just
the "condensates" introduced in ref.[1].  The formula (43) shows that due to
the "power tail" all the expectation values of $\left\langle u^{n}
\right\rangle $
 starting with $ n={3 \over2}$ are formally divergent. That simply means that
they are dominated by the non-universal region and thus change if we change
$\kappa (x) $ for $x\sim L$. At the same time, the moments with $n<{3 \over 2}$
are universal. The "power tail" in the formula (43) must be related to the
probability of having a kink, introduced in ref.[5]. However the precise
connection is not completely clear, since the non-universal part may  be
relevant in  the comparison. \par We come to the conclusion that at least in
the present setting  the violation of the naive scaling for higher  moments -
the phenomenon usualy called "intermittency"- is due to the breakdown of
Galilean symmetry and non-universality of the large velocity fluctuations. In
the past "intermittency"  essentially meant that the theory sometimes works and
sometimes doesn't. Here we have it under control.  This observation explains an
apparent discrepancy between the scaling in the eq. (41) and  Kolmogorov's
relation (19). These two come from the different regions of the phase space.
This is evident from the fact that the value of $\kappa (0)$ which enters into
Kolmogorov's relation (19) simply drops off in the G-invariant limit, as seen
from (33).    For the different type of the stirring forces considered in [5]
the two regions seem to overlap. That forms the basis for the beautiful
physical picture advocated in [5]. It  is also consistent with the approximate
solution of the Burgers problem found by the replica method in [6]. \par In the
above solution we took $a=0$. Our understanding  of other possible solutions is
still incomplete, although it seems that for  the considered type of stirring
forces an attempt to take $a\neq 0$ leads to some unphysical results, like
having  $\left\langle u(x_{1})-u(x_{2}) \right\rangle \neq 0 $. However for a
different type of  stirring forces  which lead to different scaling laws we
almost certainly   have to  include the $a$ term. This question is currently
under investigation [ 8]. \par Finally, let us present the  generalization of
the master equation for  arbitrary dimensionality. In the inviscid limit  it
isn't difficult. Consider the following quantity:
$$ \Theta_{\vec{\lambda}}=\rho (x,t) e^{\vec{\lambda}\vec{v}(x,t) } $$
where $ \rho $ and $\vec{v}$ are the density and the velocity, satisfying the
Euler equations:
$$ \dot{\rho}+\partial _{\alpha}(\rho v_{\alpha}) =0 $$
$$ \rho (\dot{v}_{\alpha}+(v_{\beta}\partial _{\beta})v_{\alpha})= f_{\alpha}
$$
It is straightforward to verify by the same methods which led to eqs. (5) and
(9) that the correlation function of  the $ \Theta _{\vec{\lambda}}$ 's :
$$ F =F (\vec{\lambda}_{1}, \vec{x}_{1}; \ldots \vec{\lambda}_{N} \vec{x}_{N})
$$
satisfies the following equation:
$$\dot{F}+\sum{{\partial^{2}F \over
\partial{\vec{\lambda_{j}}\partial{\vec{x_{j}}}}}}=\sum{\kappa_{\alpha\beta}(\vec{x}_{i}-\vec{x}_{j})\lambda_{i\alpha}\lambda_{j\beta}}F$$
which generalizes  eq.(8) for an arbitrary dimension.
 As in one dimension the origin of this equation lies in the special
conservation laws  analogous to (13). In general we have the following set of
conserved tensors:
$$ T_{\alpha_{1}\ldots \alpha_{n}}=\rho v_{\alpha_{1}}\ldots v_{\alpha_{n}},$$
which satisfy a continuity equation.
\par The next step should be an  analyses of anomalies along the same lines as
above. This task is not completed yet. Another immediate problem is to include
the pressure as a small perturbation. This is possible to do by using the
relation:
$$ \rho (\vec{x}) =\Theta_{\vec{\lambda}}(\vec{x})\mid _{\vec{\lambda}=0}$$
This relation allows us to   express the  perturbations of  pressure and
density in terms of the  function $F$. However this analyses is also a problem
for the future. \par I am deeply grateful to V. Borue for useful discussions,
A. Migdal and V.Yakhot for sharing with me  their insights, results and
enthusiasm concerning turbulence, and to D. Gross  for  his important critical
remarks on physics and style of this article.

\par This work was partially supported by the National Science Foundation under
contract PHYS-90-21984.

\eop
\appendix{REFERENCES \par [1] A.Polyakov Nucl. Phys. B396 (1993) 367 \par [2]
Ya. Zeldovich Astron. and Astroph. 5 (1972) 84\par [3]M.Kardar et al. Phys.Rev.
Lett.56 (1986) 889\par [4] Ya. Sinai Journal of Stat. Phys.84(1991) 1 \par [5]
A. Chekhlov, V.Yakhot Phys. Rev E 51(1995) R2739 and to be published\par [6] J.
Bouchaud et al. LPTENS preprint 95/12 \par [7] E. Lifschitz  L. Pitaevsky
'Physical Kinetics' Pergamon Press  (1981)  \par [8] A. Migdal A. Polyakov V.
Yakhot in preparation }

\end{document}